IAC–24–A.1.2.3

# Building Europe's first space-based Quantum Key Distribution System
## The German Aerospace Center's role in the EAGLE-1 mission

**Gabriela Calistro Rivera**[a*], **Oliver Heirich**[a], **Amita Shrestha**[a], **Agnes Ferenczi**[a], **Alexandru, Duliu**[a], **Jakob Eppinger**[a], **Bruno Femenia Castella**[a], **Christian Fuchs**[a], **Elisa Garbagnati**[a], **Douglas Laidlaw**[a], **Pia Lützen**[a], **Innocenzo De Marco**[a], **Florian Moll**[a], **Johannes Prell**[a], **Andrew Reeves**[a], **Jorge Rosano Nonay**[a], **Christian Roubal**[a], **Joana S. Torres**[a], **Matthias Wagner**[a]

[a] *German Aerospace Center (DLR), Institute of Communications and Navigation, Wessling, Germany*
<u>gabriela.calistrorivera@dlr.de</u>

\* Corresponding author

The EAGLE-1 mission aims to develop Europe's first sovereign, end-to-end space-based quantum key distribution (QKD) system. The mission is led by the European Space Agency (ESA) and SES in collaboration with several European National Space Agencies and private partners. The state-of-the-art QKD system will consist of a payload on board the EAGLE-1 low Earth orbit (LEO) satellite, optical ground stations, quantum operational networks, and key management system. The EAGLE-1 project represents a major step for next-generation quantum communication infrastructures, delivering valuable technical results and mission data, as well as contribute to the development of the EuroQCI programme.

The Institute of Communications and Navigation (IKN) of the German Aerospace Center (DLR) is a key partner in the EAGLE-1 mission and is involved in the research and development of elements in both space and ground segments. Here we report on the development of the QKD transmitter, a vital part of the QKD payload, and the customization of the Optical Ground Station Oberpfaffenhofen (OGS-OP) to conduct the IOT phase of EAGLE-1.

For the space segment, DLR-IKN is in charge of the design of the QKD transmitter, including the development of the software and firmware. This transmitter generates quantum states which are used to implement a QKD protocol based on an optical signal, that will be transmitted to ground. For the ground segment, The OGS-OP will serve as the in-orbit testing ground station for EAGLE-1. Building upon the expertise with a range of satellites for quantum communication, as well as new implementations, OGS-OP will validate the performance of the payload, optical link and QKD system for the first time. We present the main developments of OGS-OP for the mission, which includes the implementation of an upgraded adaptive optics system to correct for atmospheric distortions and optimize the coupling of the incoming light into a single mode optical fiber.

## 1. Introduction

The current quantum revolution is reshaping technologies in different areas, such as computing, sensing and communication. The advent of quantum computers introduces a potential threat for cybersecurity and security in communications as quantum algorithms could be able to easily break current widespread public and private RSA (Rivest–Shamir–Adleman) cryptographic keys. To overcome this issue, alternative methodologies need to be developed which are resistant to even the most sophisticated computing systems. Such a cryptographic system are quantum key distributions (QKD), which capitalize on properties of quantum mechanics to provide a secure communication between two users, encrypted through quantum keys. The essential property of a QKD system is that it enables the users to detect the presence of any third party trying to gain knowledge of the key, independently of the technology implemented by the eavesdropper. This property is inherent of the fundamental aspect of quantum mechanics that the process of measuring a quantum system disturbs the system, thus introducing anomalies detectable by the communicating users.

In general, quantum communication involves encoding information in quantum states, or qubits, for which photons are customary used. QKD leverage on fundamental quantum properties such as quantum superposition or quantum entanglement to establish a cryptographic key (i.e., a secret bit-string) between a pair of end-users, secured against an evesdropper. The key is considered secure if it is private, integer and authentic. This means that no third party can have any information about it (private); no third party can modify the key without being detected (integer), and the identity of the sender is verified (authentic). However, in the case the third party has access to the quantum channel, the availability of the key cannot be guaranteed (Denial-of-Service attacks, DoS)[1]. The configuration of quantum keys follow determined protocols, which can be classified in threes types: Prepare-and-Measure protocols, Entanglement-Based protocols and Measurement-Device-Independent protocols. A QKD protocol consists in both an initial quantum part and a classical post-processing sequence, which requires bi-directional classical communication (see [1] for a review). Some of the post-processing information may be encrypted and all the classical information must be authenticated, where both tasks can be achieved by using pre-shared randomness.

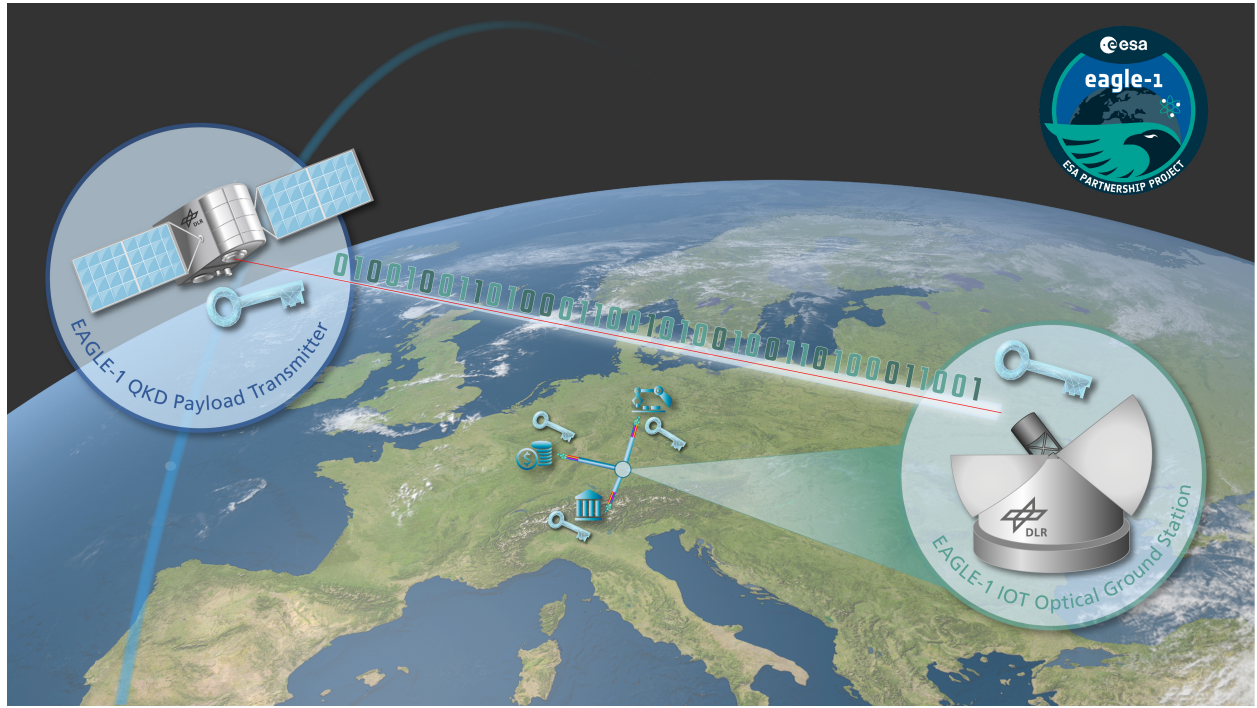

Fig. 1. The DLR-IKN participation in the EAGLE-1 mission. The DLR-IKN is a key partner in the EAGLE-1 mission and is involved in the research and development of elements in both space and ground segments.

One technological challenge for the application of QKD is that it currently relies on dedicated optic fibers as quantum channels, which impose significant geographic and technical restrictions. This challenge can be overcome with the application of QKD in space, through optical links (see reviews [2, 3] ). So far advances on this have been achieved by projects such as the satellite Micius[4] which was the first satellite which conducted QKD from space; Tiangong-2[5], where the Chinese space station hosted a compact QKD payload performing downlink to several ground stations; and Jinan 1 [6] which followed upon the Micius satellite with a reduced size.

Developing space QKD for Europe is crucial to achieve secure and autonomous communication. At the forefront of this development is the EAGLE-1 QKD system[7], which will demonstrate and validate QKD technologies and provide Europe with the first end-to-end QKD system, paving the way for ultra-secure data transmissions within Europe. EAGLE-1 is an ESA Public Private Partnership project, co-funded by the ESA contribution of Germany, Luxembourg, Austria, Italy, the Netherlands, Switzerland, Belgium and the Czech Republic under the ESA's ARTES Scy-Light (SeCure and Laser communication Technology) programme, as well as the European Commission through Horizon Europe. EAGLE-1 is key to the European Quantum Communication Infrastructure (EuroQCI), which includes 24 EU Member states and develops both space and ground solutions for QKD. The EAGLE-1 satellite is expected to be launched with the Vega C rocket in late 2025 to early 2026.

It will then complete three years of in-orbit validation supported by the European Commission, with an option to extend the mission by further two years.

The EAGLE-1 system is designed, developed and operated by a consortium of 20 European partners, led by SES, with ESA and European Commission support. The Institute for Communications and Navigation (IKN) of the German Aerospace Center (Deutsches Zentrum für Luft und Raumfahrt, DLR) is a key member of this consortium, contributing to both the space and the ground segments of the project (see Figure 1). In particular, DLR-IKN is in charge of the design and development of the QKD transmitter which will be implemented within the QKD Payload of the EAGLE-1 satellite. Furthermore, DLR-IKN develops and applies the in-orbit testing (IoT) Ground Station for EAGLE-1 which is hosted at the institute's premises. The DLR-IKN contribution is supported by the institute's involvement in the predecessor Quantum Crytography Telecommunication System (QUARTZ) project, which provides the baseline for the design and development of the QKD transmitter, as well as by the institute's expertise on optical satellite links with DLR's Optical Ground Station Oberpfaffehofen (OGS-OP).

## 2. Space segment: the QKD Transmitter

DLR-IKN contributes to the space segment of EAGLE-1 with the design of the QKD Transmitter. The design includes the electrical and optical design, as well as the final development of the software and FPGA. In collaboration with the consortium partner TESAT, we will integrate a fully functional QKD transmitter into the satellite QKD payload.

The main task of the QKD Transmitter is to convert protocol information from the QKD Processor into an optical signal in the form of encoded single photons. The modulated photons (quantum key bits) are then distributed through an optical terminal from the satellite to the optical ground station (see Section 3). The QKD protocol implemented in EAGLE-1 is based on Bennett and Brassard 1984 [BB84, 8], which features the best trade-off between implementation security and system performance [1].

The QKD transmitter is a complex system that consists of electrical modules and optical modules, driven by a microcontroller software and FPGA design.

### 2.1 Electronic modules

The electronic modules of the QKD Transmitter are shown in the left side of Figure 2. The transmitter is connected to the QKD power unit and to the QKD Processor. As input, the transmitter receives key and protocol information from the QKD processor that contains a quantum random number generator (QRNG). The electronic high-speed board contains the FPGA (Field Programmable Gate Array), power converters, communication interfaces, memories, clock tree and the electro-optical interfaces.

*2.1.1 FPGA*

The FPGA serves for two important functions: the generation of modulation signals and the control & monitoring of the transmitter.

To fulfill the first task tthe FPGA transforms the high-speed input data from the QKD Processor into electro-optical signals for the modulators following the EAGLE-1 QKD protocol. Therefore, the FPGA tranceivers drive high-speed digital-to-analogue-converters (DACs) which in turn drive the modulators. The final modulation (qubit) rate for the key is 2.25 Gbps.

The second FPGA functionality is control and monitoring. The FPGA hosts an embedded microprocessor core that runs the software of the QKD transmitter (see Section 2.3 for details on the software). Additionally, the FPGA implements microcontroller interfaces of the hardware peripherals and a communication interface for telemetry and telecommands.

*2.1.2 Electro-optical interfaces*

The electro-optical interfaces consist of high-speed DACs, low-speed DACs and analogue-to-digital converters (ADCs) for optical measurements. The high-speed DACs are used to generate the modulation signals including amplitude and phase. The amplification and attenuation require the control of power level, as well as feedback measurements of optical power levels. The modulators also require a control of several biases.

### 2.2 Optical modules

The main optical modules of the transmitter unit are shown in the right side of Figure 2. The optical modules consist of five blocks: pulse generation, including optical modulation; the calibration unit; pulse amplification; pulse attenuation; and a quantum hacking protection.

The pulse generation module creates the optical signal. A C-band laser creates the light that is needed to pass through the optical setup. The optical modulators create pulses with individual phases for each pulse according to the EAGLE-1 QKD protocol.

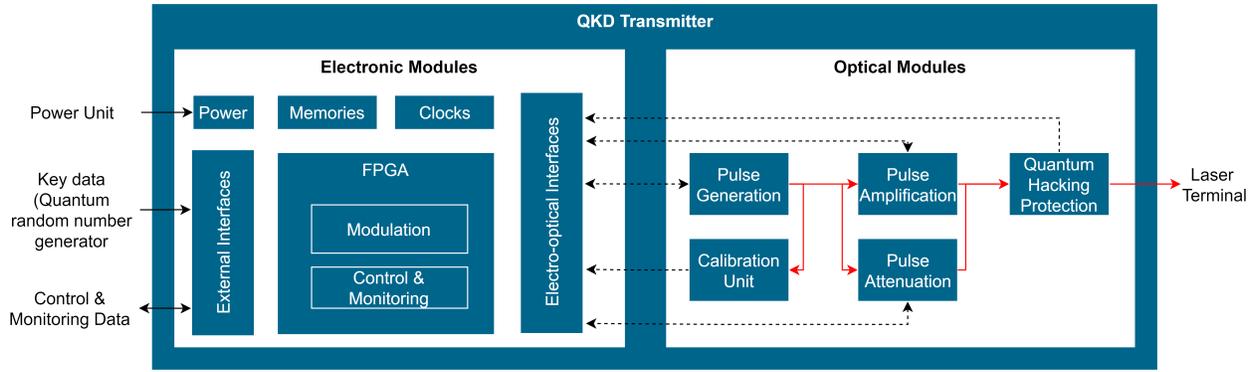

Fig. 2. Block diagram of the main modules of the QKD Transmitter.

The calibration unit verifies the phase differences between the pulses. The transmitter unit requires a calibration for initial start-up and operation across temperature variations. This calibration is based on feedback control with several photodiodes for correct amplitudes, biases and timings.

After the pulse generation, the pulses are either amplified to be used as reference signals or attenuated to single photon level for the key bits. In particular, the pulse amplification modules include a semiconductor optical amplifier (SOA) and a variable optical attenuator (VOA). The SOA acts as an optical switch, which attenuates light when it is turned off and amplifies it when it is turned on. The VOA is used to control the output power for reference and quantum pulses so that these achieve the required mean power at the output. The pulse attenuation part controls the transmitter output to single photon levels per key bit.

Finally, the hacking protection contains countermeasures for external attacks. The optical modules output then the final signals into the laser terminal which will downlink to the ground station (Section 3).

2.3 QKD Transmitter Software

The software is configured in the microcontroller core hosted in the FPGA (Section 2.1.1). The main functions of the software are calibration of the QKD Transmitter, the coordination of telemetry and telecommand, monitoring of the hardware, hacking detection, and maintaining a running system, such as running diagnostics, software updates and data management.

Calibration is required as environment variables on the satellite change during operation in space. There is a constant need to ensure that the QKD Transmitter operates within a parameter regime which enables the generation of a secret key. Degradations in output signals due to changes in the environmental variables must therefore be compensated continuously.

Sensors located in the QKD Transmitter continuously generate measurement values, that are processed and monitored by the QKD Transmitter software. When the QKD Transmitter is in operation, these measurement values give information about the health of the hardware in particular and the QKD Transmitter in general.

Communication with the transmitter in orbit is a further necessary functionality. This task is facilitated by the software component related to telemetry and telecommand to the ground.

The software is developed according to ECSS. Automatic testing and linting in a CI/CD (continuous integration and continuous deployment) pipeline and a test-driven development paradigm ensure compliance with the standard and enable quick and sturdy development. Software development adheres to clean code, clean architecture guidelines and SOLID principles.

2.4 The QKD Transmitter electrical and functional model (EFM)

In Fig. 3 we show the QKD Transmitter EFM (Electrical and Functional Model) that has been designed and built by DLR-IKN. The EFM is functionally representative of the end product in electrical, optical and software terms. It is being used for functional and interface tests and for failure mode investigations.

Our EFM consists in a multilayered set-up, which has a flexible setup and structure that allows the active research and development of the final design. As described in Fig. 3, the two central layers corre-spond to the main modules of the QKD Transmitter. These include the main board with the FPGA, micro-controller and software, electro-optical interfaces and associated active optical components (modulators for pulse generation and pinboard for monitoring/calibration), which are located at the top layer. The lower layer include all passive optical components, which are connected to the upper components via the connector matrix, visible in the upper layer. Both drawers are linked and are pulled out together to form the QKD transmitter, as described in Fig. 2.

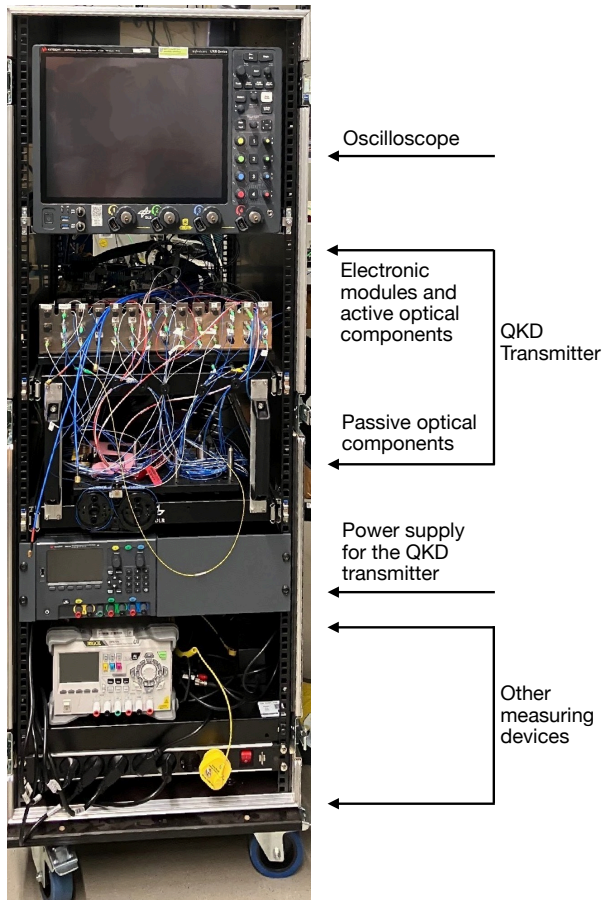

Fig. 3. QKD Transmitter Electrical and Functional Model (EFM), which is being currently deployed by DLR-IKN.

The rest of the rack provides space for external equipment such as the oscilloscope for measuring and debugging electronic and optical signals (upper panel), power supply for transmitters (black under the drawers), and other measuring devices such as photo-receiver, coherent receiver, and the optical power meter.

The mobile setup inside a rack allows continuous consorted experiments with the QKD receiver, which is being developed in parallel by consortium partners. This EFM model defines and validates the final design and represents the technical foundation of the system, preceding the qualification and flight models built in collaboration with our industry partner TESAT.

## 3. Ground segment: The IoT Optical Ground Station

The contribution of DLR-IKN to the ground segment of EAGLE-1 builds upon the institute's expertise in optical links for classic and quantum optical communication, as well as two decades of experience in the development of optical ground stations.

### 3.1 DLR-IKN Optical Ground Station Oberpfaffenhofen

DLR-IKN hosts on its rooftop the Optical Ground Station Oberpfaffenhofen (OGS-OP, Figure 4). OGS-OP has performed a range of successful satellite and aircraft measurements with the satellites OICETS [9], SOCRATES[10], ISS and the DLR aircraft Do228, as well as a flight experiment where the world first quantum communication between a flying carrier and a ground station was demonstrated [11]. These experiments were conducted with a 40 cm Cassegrain telescope which is now uninstalled.

Recently, a significant upgrade (OGS-OP Next Generation; OGS-OP NG) has been developed with the installation of a new 80 cm telescope in summer 2021 (Figure 4). This new telescope is currently employed in bidirectional links with adaptive optics (AO) and AO-predistortion to geostationary satellites, and for the checkout of the various LEO satellites carrying diverse OSIRIS-terminals (Optical Space InfraRed link System)[12]. The station's multiple focii, including the Coudé laboratory shown in Figure 5, provides the OGS-OP NG with high flexibility to enable multi-mission support. This telescope together with an existing optical bench setup and lab infrastructure form the basis for the further development of the OGS-OP NG to become the EAGLE-1 IoT OGS.

### 3.2 The EAGLE-1 IoT OGS

Building upon recent improvements of OGS-OP NG, DLR-IKN develops and runs the In-Orbit-Testing Optical Ground Station for the EAGLE-1 mission (IOT OGS) in Oberpfaffenhofen (Figure 4). The IOT OGS is designed to conduct scientific measurement campaigns to commission the QKD payload during the first 6 months of its mission.

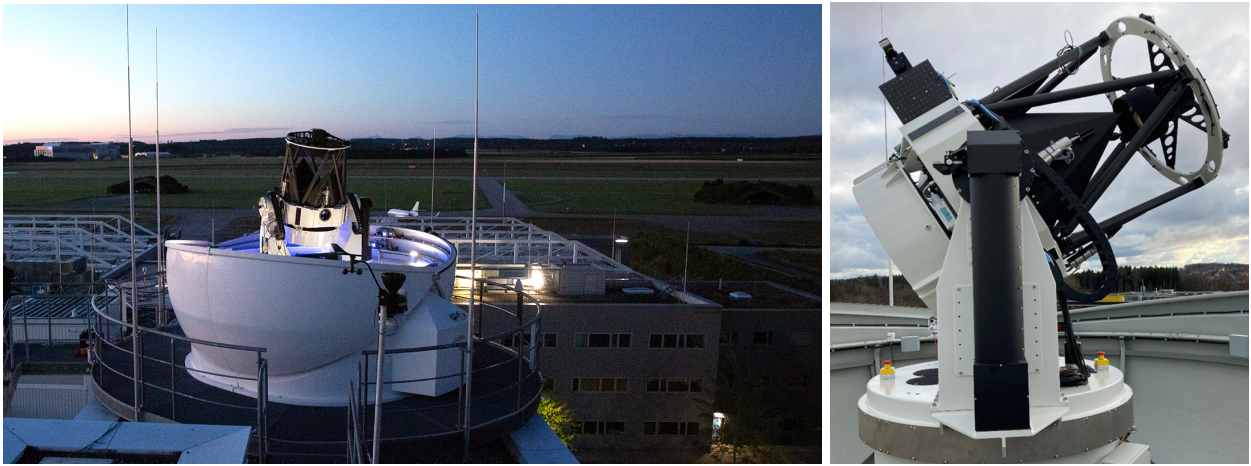

Fig. 4. The left panel shows the OGS-OP NG on the rooftop of DLR-IKN in Oberpfaffenhofen, southern Bavaria, Germany. The right panel zooms in onto the 80 cm Nasmyth-Design telescope.

With its 80 cm primary mirror, the IOT OGS is designed to track the satellite on LEO and communicate with the QKD payload via optical communication links, supporting the system QKD wavelengths as well as uplink and downlink signals. To optimize the link budget required for the QKD system, it is furthermore crucial to mitigate signal losses caused by atmospheric turbulence. The IOT OGS thus includes a tip-tilt mirror and a high-order Deformable Mirror (DM) as part of an advanced Adaptive Optic System, which will be described below (Section 3.2.3). A block diagram in Figure 6 shows the IOT OGS capabilities already provided by the OGS-OP NG (white blocks), as well as the on-going and planned upgrades that are tailored for the EAGLE-1 system (orange blocks). We will expand on these components below.

3.2.1 Pointing, Aqcuisition and Tracking

In order to establish communication between two optical terminals, the communication terminals are required to perform spatial acquisition with the goal of constant tracking. Spatial acquisition is the process of coaligning the shared line of sight between the transmitter and receiver. First, acquisition needs to take place, ideally based on accurate orbit files and pointing accuracy of the ground stations, or otherwise a lead-and-follow acquisition-scheme. Upon detection of the uplink communication laser system, the satellite optical terminal begins closed-loop tracking and then proceeds with data transmission. The OGS then receives the downlink and uses this data signal to track the satellite in closed-loop. If the link is interrupted, e.g. by clouds, re-acquisition is attempted.

3.2.2 Transmitting and receiving channels

Figure 6 shows the main components of the uplink communication and the uplink path as red dashed lines. The transmitting channel in EAGLE-1 consists of an uplink beacon system of 4 laser beams (U1-4) used to mitigate scintillation at receiver. The beam collimators are mounted directly on the telescope assembly co-aligned to the telescope aperture. The transmitted light goes directly through the collimators, which guarantees high channel isolation, i.e. no back reflection or additional losses by the Coudé optics, with moderate system complexity. Each of the four transmitting beacon laser is shifted by frequency in order to achieve transmitter diversity and avoid far-field interference for mitigation of scintillation.

The free space recieving channel is depicted in Figure 6, where the main components and the downlink path are shown as green solid lines and arrows. The first and core element of the receiving channel is the Nasmyth-Design Ritchey-Chretien telescope of 80 cm main aperture diameter. Its Nasmyth-Design allows the light to be directed through the Coudé-Path onto an optical table in a laboratory room below the telescope mount, wherein wavefront corrections are performed (AO; Section 3.2.3) before the light is coupled into a single mode fibre. A new chromatic coupling system is developed and integrated to comply with the EAGLE-1 wavelength plan. The optical table is shown in Figure 5. The optical design of the telescope covers a Nasmyth-port, a Coudé port, and two folded Cassegrian ports, where the coudé port is used by the EAGLE-1 setup. [13].

### 3.2.3 The adaptive optics system

To optimize the coupling between the received downlink signal and the single mode optical fibre, an adaptive optics (AO) system is required to correct for atmospheric turbulence which imprints the wavefronts that enter the telescope. Atmospheric turbulence is particularly strong at satellite observation at low elevation angles and fast-changing conditions, i.e. due to winds. These turbulence degradations can then be mitigated via the following system (as shown in Figure 6). As soon as the laser light from the satellite terminal appears as a constant light source within the acquisition camera, closed-loop tracking can be engaged such that the channel is on-axis with the Coudé bench. Turbulence degradations are measured using a Shack-Hartmann wavefront sensor (SH-WFS), which drives a fast steering (tip-tilt) mirror and a high-order Deformable Mirror (DM) in closed-loop. The update rate for the high-order wavefront corrections is on the order of multiple kHz. The main purpose of including the closed-loop adaptive optics system is to optimise the single mode fibre coupling efficiency. The communication signal is then guided to the receiver front ends which converts the optical signal to electrical signal and will be transmitted to the modem for data reception.

### 4. Conclusions

The EAGLE-1 mission will demonstrate Quantum Key Distribution from LEO. As part of a consortium of more than 20 EU Member states, DLR-IKN develops the design, software and firmware of the QKD transmitter, the IOT OGS and participates in the IOT. The QKD transmitter implements a BB84 protocol in optical C band. The OGS-OP is modified to serve as OGS IOT in the EAGLE-1 mission. Both systems, QKD transmitter and IOT OGS, are currently in development. The planned satellite launch is between end of 2025 and early 2026. When in orbit, the EAGLE-1 satellite will be the first European QKD satellite and enable extensive testing of the QKD system and the QKD service.

### 5. Acknowledgements

We thank Thorsten Schubert and our colleagues from ANDAV Electronics GmbH and Schmidt Embedded Systems for their significant contributions as external members of the team. We acknowledge the collaboration with Tesat in the development of the QKD Transmitter and its integration into the QKD payload, as well as the cooperation with FAU on the consorted testing with the QKD receiver. This work was supported by funds from the Federal Ministry for Economic Affairs and Climate Action of Germany (BMWK), which were provided through the ESA program SAGA. We acknowledge the cooperation and support from SES, ESA and the European Commission.

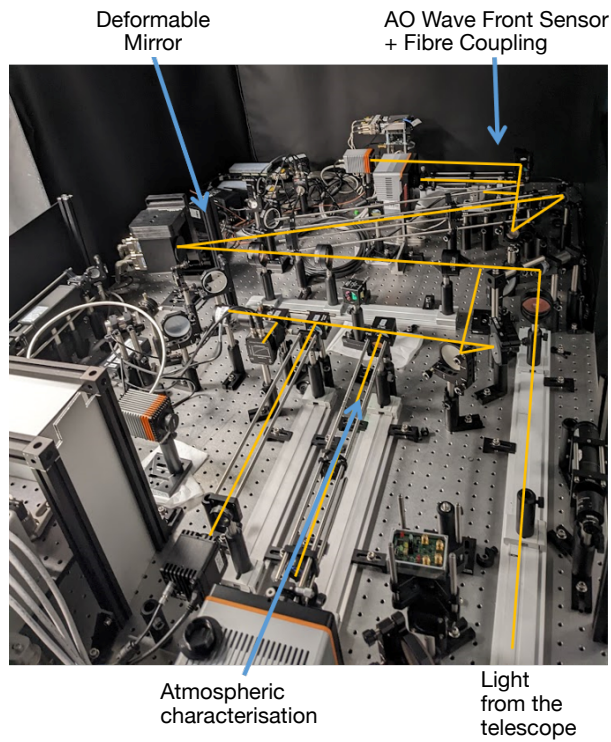

Fig. 5. Coudé laboratory setup at the EAGLE-1 IOT OGS.

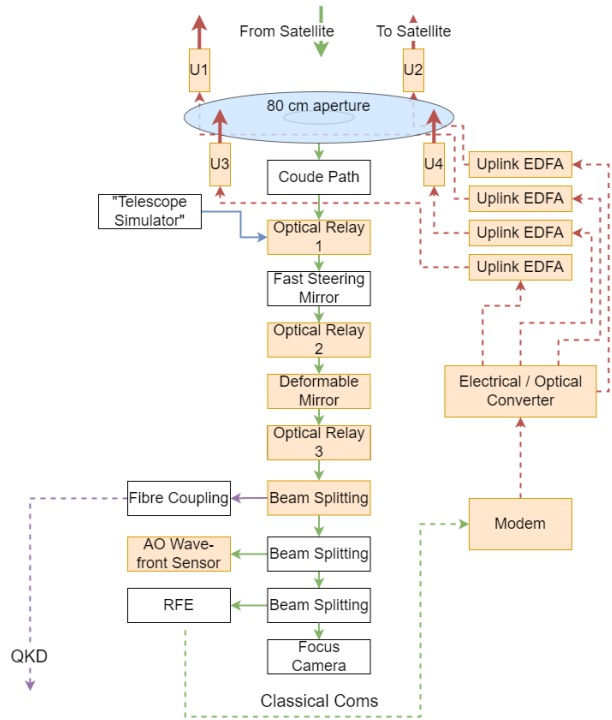

Fig. 6. A Block diagram of the proposed receiving and transmitting setup of the OGS-OP. Orange blocks mark changes made within EAGLE-1. Red dashed lines mark the fiber and electrical uplink path. Purple dashed line mark the fiber coupled QKD path. Green solid lines show the free space receiving downlink path and the green dashed line depicts the electronic signal between the Receiver Front End (RFE) and the Modem.